\documentclass[twoside,12pt]{article}
\usepackage{epsfig}

\def\Journal#1#2#3#4{{#1} {#2} (#4) #3 }

\def\NPA{{\em Nucl. Phys.} A}

\def\PLB{{\em Phys. Lett.} B}

\def\PRL{\em Phys. Rev. Lett.}

\def\PREP{\em Phys. Rep.}

\def\PRD{{\em Phys. Rev.} D}
\def\PRC{{\em Phys. Rev.} C}

\newcommand{\be}{\begin{equation}}
\newcommand{\ee}{\end{equation}}
\newcommand{\bea}{\begin{eqnarray}}
\newcommand{\eea}{\end{eqnarray}}

\begin{document}

\title{ \vspace{1cm} On the accelerated expansion of the cosmos}
\author{Dominik J. Schwarz,\footnote{Talk given at the School on Nuclear Astrophysics, Erice 2010}, 
Benedict Kalus and Marina Seikel \\
Fakult\"at f\"ur Physik, Universit\"at Bielefeld, Postfach 100151, 33501 Bielefeld, Germany}
\maketitle
\begin{abstract} 
We present a short (and necessarily incomplete) review of the evidence for the accelerated expansion 
of the Universe. The most direct probe of acceleration relies on the detailed study of supernovae (SN) 
of type Ia. Assuming that these are standardizable candles and that they fairly sample a 
homogeneous and isotropic Universe, the evidence for acceleration can be tested in a model-
and calibration-independent way. Various light-curve fitting procedures have been proposed 
and tested. While several fitters give consistent results for the so-called Constitution set, they lead to 
inconsistent results for the recently released SDSS SN. Adopting the SALT fitter and relying 
on the Union set, cosmic acceleration is detected by a purely kinematic test at $7\sigma$ when spatial flatness is assumed and at $4\sigma$ without assumption on the spatial geometry. A weak point 
of the described method is the local set of SN (at $z < 0.2$), as these SN are essential to anchor the 
Hubble diagram. These SN are drawn from a volume much smaller than the Hubble volume 
and could be affected by local structure. Without the assumption of homogeneity, 
there is no evidence for acceleration, as the effects of acceleration are degenerate with the effects of inhomogeneities. Unless we sit in the centre of the Universe, such inhomogeneities can be constrained by SN observations by means of tests of the isotropy of the Hubble flow. 
\end{abstract}
%\eject
%\tableofcontents
\section{Introduction}

Until the early 1990s it was widely believed that the Einstein-de Sitter model would be 
an appropriate description of the Universe. This believe was built on the observational evidence 
for the existence of cold dark matter and the idea of cosmological inflation, predicting the spatial 
flatness of the Universe. The Einstein-de Sitter (EdS) model predicts that the Universe decelerates with 
a constant deceleration parameter $q = 1/2$. In 1993 this value was still consistent with measurements 
of the angular diameter distance of a certain type of radio galaxies \cite{k93}. The EdS model also makes a definite prediction for the age of the Universe $t_{\rm EdS} = 2/(3 H_0) $, from a measurement of todays Hubble expansion rate $H_0$. In the middle of the 1990's it became clear that $H_0$ is 
not small enough to allow for an age of the Universe that exceeds the age of the oldest 
objects observed. This led to an ``age crisis'' and prepared the ground 
for the cosmological constant to reenter the game \cite{bh95,os95}.

At the turn of the millenium the revolution finally took place. By means of the observation of supernovae of type Ia, which are believed to provide the best standardizable and extremely bright candles in the 
Universe, the EdS model was ruled out at a statistical significance of about $3\sigma$ by two independent groups \cite{rea98,pea99} and it was demonstrated that adding a cosmological 
constant allows to fit the supernova data and to solve the age crisis at the same time. Although this 
result was welcomed with some reservation by the community, soon it was backed up by the
detection of the first accoustic peak in the angular power spectrum of the temperatur anisotropies of the cosmic microwave background (CMB) \cite{mea99,dbea00,hea00}. The CMB, together with a measurement of $H_0$, established that the spatial geometry of the Universe is euclidian and 
confirmed this generic prediction of cosmological inflation. The new model was called concordance
model, because it fitted the CMB and SN Ia besides other probes. The determination of the cold dark matter to baryon ratio from galaxy clusters, together with the baryon density obtained from primordial nucleosynthesis and the observation of light element abundances, provided a measurement of the 
total matter density. It turned out to be in concordance with the value found by the combination of 
CMB and SN Ia, a first nontrivial test of the new model. 

Largely unnoticed by the majority of the cosmology community it was mentioned already in 2000 that 
this interpretation of the data relies heavily on the assumption of homogeneity of the Universe 
\cite{c00}. It was shown that an inhomogeneous Universe does fit the SN Ia Hubble diagram. 
Already earlier inhomogeneous models had been proposed to resolve the age crisis. Nevertheless, 
this interpretation of the observations has two big disadvantages. First of all, inhomogeneous models are less predictive, as there is no proposal for appropriate initial conditions 
(in the concordance model cosmological inflation predicts the isotropy and homogeneity of a region 
of the size of at least a Hubble volume today), secondly although possible in principle, no precise predictions for the CMB anisotropies have been calculated and thus the CMB cannot be used to test those models, apart from qualitative tests relying on the position of the first acoustic peak. Thus, this 
was not considered as a serious alternative and the $\Lambda$ cold dark matter model ($\Lambda$CDM) became the new standard.

The concordance model is a minimal model that relies on \\
1.~established particle physics, encoded by the temperature of the CMB $T_0$ and the 
dimensionless energy density parameters of baryonic matter $\Omega_{\rm b}$ (and neutrinos
$\Omega_\nu$), \\
2.~the theory of general relativity with a cosmological constant, introducing the Hubble parameter 
$H_0$ and the cosmological constant $\Lambda$, \\
3.~the idea of comological inflation, which includes the assumptions of isotropy and homogeity 
and predicts spatial flatness and gaussian, scale-invariant, and isentropic fluctuations. 
Its physics is parametrized by the amplitude of density perturbations $A$, the spectral tilt of their 
power spectrum $n-1$, and the tensor-to-scalar ratio $r$, \\
4.~the existence of dark matter $\Omega_{\rm cdm} = 1 - \Omega_{\rm b} - \Omega_\Lambda$, \\
and astrophysical parameters that encode complex physics like the optical depth $\tau$, bias $b$, 
the calibration of SN Ia ${\cal M}, \dots$

When fitting this model to the observations of the Wilkinson Microwave Anisotropy Probe (WMAP)
\cite{lea10}, a modern measurement of $H_0 = 74.2\pm 3.6$ km/s/Mpc \cite{rea09} and baryon 
acoustic oscillations in the distribution of the large scale structure \cite{rea10,pea10}, one finds 
that about  $72$\% of the cosmic energy budget is made up of the cosmological constant (or dark energy), $23$\% is contributed by dark matter and only $5$\% is contributed by atoms and other 
forms of known matter \cite{lea10}. 

As these numbers are not less and not more than just a fit to a particular model, the mere 
existence of the fit is not a proof of the accelerated cosmological expansion. A direct proof of 
accelerated expansion is of utmost importance for cosmology. For isotropic and homogeneous cosmology, we can rewrite Einsteins equation as 
\begin{equation} 
- 3 \frac{\ddot{a}}{a} = 4 \pi G (\epsilon + 3 p),   
\end{equation} 
where $a$ denotes the scale factor, $\epsilon$ is the energy density and $p$ denotes the pressure.  
Thus establishing $\ddot{a} > 0$ in a model independent way would either proof the existence of 
dark energy with $p < - \epsilon/3$, a violation of Einsteins equation, or a violation of the 
assumptions of isotropy and homogeneity. 

\section{Model- and calibration-independent test}

These considerations motivate a model-independent (or as model independent as possible) test or measurement of $q \equiv - (\ddot{a}/a)/H^2$. One possibility would be to just Taylor expand the 
Hubble law (which is a direct consequence of the asumptions of homogeneity and isotropy). The 
coefficient of the term quadratic in redshift z would allow us to measure $q_0$. However, the problem 
here is that the radius of convergence of a Taylor series in $z$ is
$<1$ \cite{cat07} and thus already at $z=0.3$ 
there are non-negligible corrections from higher orders. 
Nevertheless, such Taylor expansions of different quantities have
been done \cite{vis05,elg06}.

Also other parametric and non-parametric attempts to reconstruct $q$
\cite{rap07,st06} or $w\equiv p/\epsilon$ \cite{cz10,hol10}
have been put forward. Although these approaches do not use a
specific cosmological model, they all rely on certain
assumptions. They assume the validity of Einstein's equation and
spatial flatness or assume some arbitrary evolution of $q$.

However, if one is not interested in the exact reconstruction of
$q$, but in the question whether there has been a phase of cosmic
acceleration, there is a much simpler test \cite{ss08,ss09}.
One still needs to assume homogeneity and isotropy, but can 
drop all other assumptions (matter content, model of dark energy, Einstein equation, spatial curvature).
It is the test of the null hypothesis $q \geq 0$ for all redshifts. If we can reject that hypothesis, we can proof that there was an epoch of accelerated expansion. The inequality $q \geq 0$ translates into an 
inequality for the distance modulus $\Delta \mu \equiv \mu - \mu(q=0) \leq 0$. The distance modulus is 
a logarithmic measure of the brightness of a SN. There is still a problem. In order to perform this test, we would need to know what the absolute magnitude (luminosity) of SN Ia is. In order to get rid of this 
calibration issue, we look only at $\Delta \mu - \Delta \mu_{\rm nearby}$, which means that we use the nearby set of SN to self-calibrate the test \cite{ss09}.

In previous work \cite{ss09}, we applied this test on the Union set \cite{kea08}. 
The null hypothesis of no acceleration was rejected at a statistically significant level 
($>7\sigma$ if spatial flatness was assumed, $> 4\sigma$ if 
arbitrary spatial curvature was allowed). 

\begin{figure}[h]
\begin{center}
\begin{minipage}[t]{8 cm}
\epsfig{file=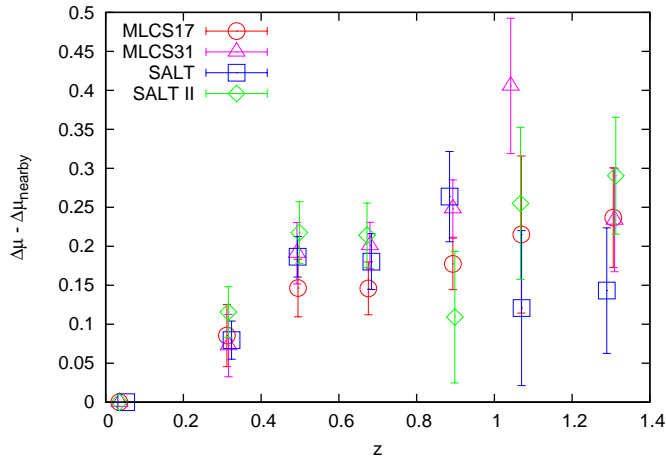,scale=1}
\end{minipage}
\begin{minipage}[t]{16.5 cm}
\caption{Evidence for accelerated cosmic expansion from the
  Constitution set \cite{hea09} assuming spatial flatness. 
This set of SN Ia was analyzed with four different light-curve fitters. All of them give rise to 
consistent results and reject the null hypothesis of no acceleration at an statistically significant level. \label{fig1}}
\end{minipage}
\end{center}
\end{figure}

\begin{table}[t]
\begin{center}
\begin{tabular}{lllll}
\hline
                    & MLCS17      & MLCS31      & SALT        & SALT II\\
\hline
spatially flat      & $6.5\sigma$ & $7.9\sigma$ & $8.2\sigma$ & $7.6\sigma$\\
arbitrary curvature & $2.5\sigma$ & $4.3\sigma$ & $4.7\sigma$ & $4.2\sigma$\\
\hline
\end{tabular}
\begin{minipage}[t]{16.5 cm}
\caption{\label{constitution-tab}Evidence for accelerated cosmic expansion from the
  Constitution set \cite{hea09} for the four light-curve fitters.}
\end{minipage}
\end{center}
\end{table}

More recently we updated \cite{s10} 
our analysis for the Constitution set \cite{hea09}, as this data set fills in a 
large number of nearby SN Ia, which play an important role in our test, as they are the anchor of the calibration. Moreover, the Constitution set was analyzed for four light-curve fitters, which allows us to study systematics. The result is consistent with the analysis of the Union sample and is shown in 
Figure 1. The evidence for acceleration is given in table
\ref{constitution-tab}. Assuming a spatially flat universe, it lies
between $6.5\sigma$ and $8.2\sigma$ for the different light-curve
fitters. Allowing for an arbitrary spatial curvature, the evidence is
significantly reduced, but (with exception of MLCS17) is still larger
than $4\sigma$.

\begin{figure}[h]
\begin{center}
\begin{minipage}[t]{8 cm}
\epsfig{file=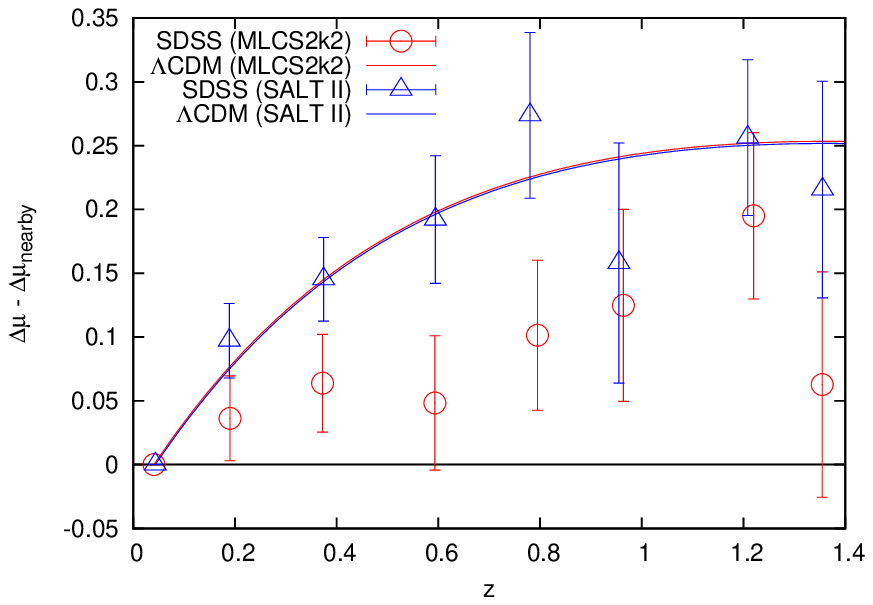,scale=1}
\epsfig{file=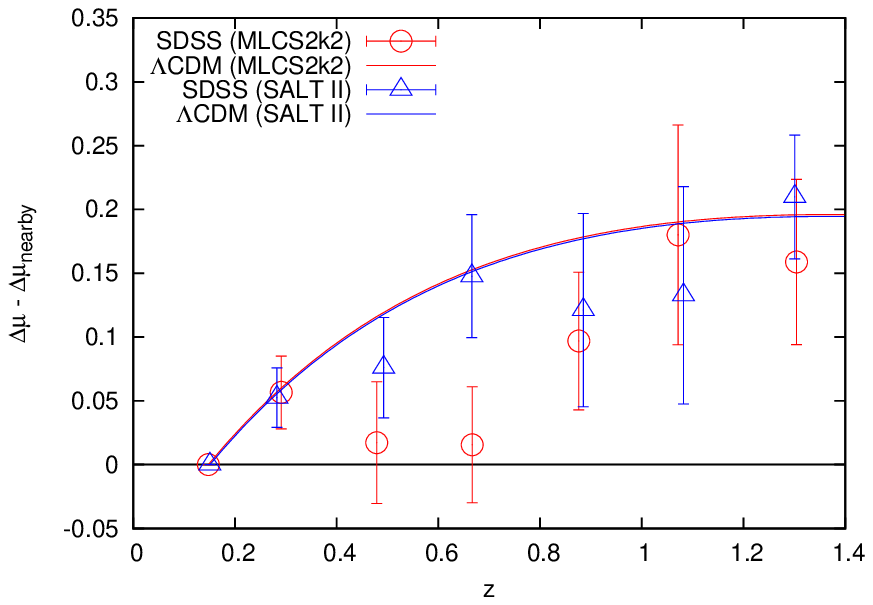,scale=0.75}\epsfig{file=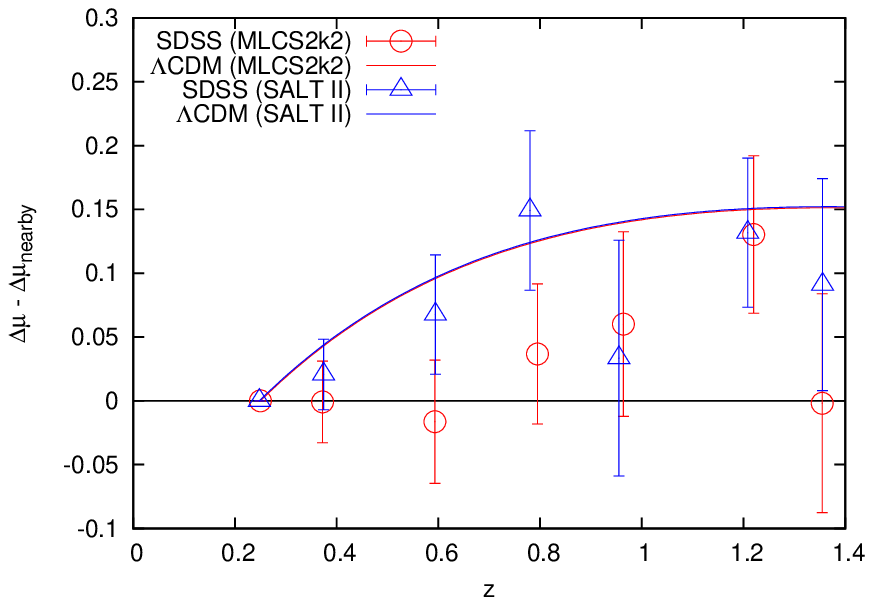,scale=0.75}
\end{minipage}
\begin{minipage}[t]{16.5 cm}
\caption{The case for accelerated expansion, based on the SDSS SN \cite{kea10}, is not convincing. The two light-curve fitters give results that differ significantly. In the top panel, the reference bin contains 
SN below $z =0.1$. Below, SN from $z = 0.1$ to $0.2$ (left) and $z = 0.2$ to $0.3$ (right) are used for self-calibration. In all panels, spatial flatness is assumed. \label{fig1}}
\end{minipage}
\end{center}
\end{figure}

As both the Union and the Constitution samples contain very few SN Ia at intermediate redshifts 
($z = 0.1$ to $0.3$) the Sloan Digital Sky Survey SN \cite{kea10} have been awaited with great 
interest. Consequently we applied our test to this sample as well and found a result that does not fit 
into the harmonic picture that unravelled so far \cite{s10}. 
The surprise comes when looking again at the two 
different light curve fitters used (MLCS2k2 and SALT II). This is shown in the top panel of Figure 2. 
It turns out that the results are just inconsistent with each other. It was indeed noticed in \cite{kea10}
that the two results do not agree, but when marginalizing over the calibration the strong 
discrepancy between the light-curve fitters is not as obvious as our case. Here we anchor the 
calibration at a particular bin and cannot hide any difference in a nuisance parameter. 
We also checked whether this situation changes if nearby SN are dropped. The reason for that is that 
the nearby sample of SN must be most sensitive to inhomogeneities. However, even rejecting all SN 
at $z<0.2$ does not resolve the conflict between the light-curve fitters.  

We conclude that the published SDSS SN data suffer from some severe
issue linked to one or both of the light-curve fitters. While using
the SALT II fitter  the SDSS SN analysis is consistent with findings 
from the Union and Constitution compilations, the analysis based on MLCS2k2 is not --- the null hypothesis is not rejected at a statistically significant level.

\section{Testing of the underlying assumptions}

Although the SDSS SN analysis is inconclusive, we could argue that the accelerated cosmic 
expansion is well established, based on the Union and Constitution samples.  Apart from 
systematic issues in the observation and analysis of SN Ia data, we also have to test 
our fundamental assumptions --- homogeneity and isotropy. 

The calibration-independence of our test as presented above, is achieved by anchor all SN wrt 
a sample of nearby SN. However, the nearby sample is collected at  $z\ll 1$, e.g.~in Figure 1
at $z<0.1$. Thus this means that we have implicitly assumed that a volume of $\sim (400 {\rm Mpc})^3$ 
is a good and fair representation of the Hubble volume. Is that justified? This is not obvious. On the one hand the homogeneity scale is believed to be $\sim 100$ Mpc. On the other hand the largest known 
structure in the Universe is the Sloan Great Wall, extending over $\sim 400$ Mpc.  

One way of avoid these ``small'' scales would be to use only SN 1a at $z> 0.1$. An attempt  for such a
test is also shown in Figure 2 in case of the SDSS SN (it does not
work for the Constitution set, as it contains only few SN at $z = 0.1$
to $0.3$). Table \ref{sdss-tab} lists the evidences for acceleration
for different choices of the set of nearby SN, which are used for the
calibration. 
If we focus on the SALT analysis, we see that the evidence
for acceleration reduces if we go to higher redshift, however the advantage is that the ``nearby'' SN sample larger volumes and the assumption of homogeneity and isotropy is probably justified 
(at least in the context of the concordance model).  
It will be interesting to repeat this analysis with the 
3yr SNLS data (see talk by R. Pain at this school) and the SN Factory (see talk by M. Kerschhaggel
at this school).

\begin{table}[t]
\begin{center}
\begin{tabular}{lll}
\hline
calibration bin & MLCS2k2     & SALT II \\
\hline
$0.0\le z< 0.1$ & $1.7\sigma$ & $4.5\sigma$\\
$0.1\le z< 0.2$ & $2.0\sigma$ & $3.5\sigma$\\
$0.2\le z< 0.3$ & $0.3\sigma$ & $2.0\sigma$\\
\hline
\end{tabular}
\begin{minipage}[t]{16.5 cm}
\caption{\label{sdss-tab}Evidence for accelerated cosmic expansion from
  SDSS SN \cite{kea10} for the two light-curve fitters and different
  calibration bins.}
\end{minipage}
\end{center}
\end{table}

Instead of rejecting the very local sample of SN, 
we could also use the SN Ia data to study the isotropy and homogeneity of the 
cosmic expansion.  It is well known that Lemaitre-Tolman-Bondi (LTB) models, which are sperically symmetric, but inhomogeneous, can fit almost any Hubble diagram, 
as this model contains two free functions of $r$. However, the LTB model violates the Copernican principle most strongly --- we would live in the centre of the Universe. Back to the middle ages?
However, we could view the LTB model as a toy model for our local neighborhood and imagine that there are many such almost spherical voids in  the Universe (see also to talk of T.~Cliffton at this 
school). 
 
Below we to adopt the point of view that our local neighborhood might be inhomogeneous and anisotropic. Even if it is an almost shperical over or underdensity, it is very unlikely that we live at 
a local center. Any dislocation from a center of a void or big supercluster would lead to a sizeable 
anisotropy in the local expansion. 

\begin{figure}[h]
\begin{center}
\begin{minipage}[t]{8 cm}
\epsfig{file=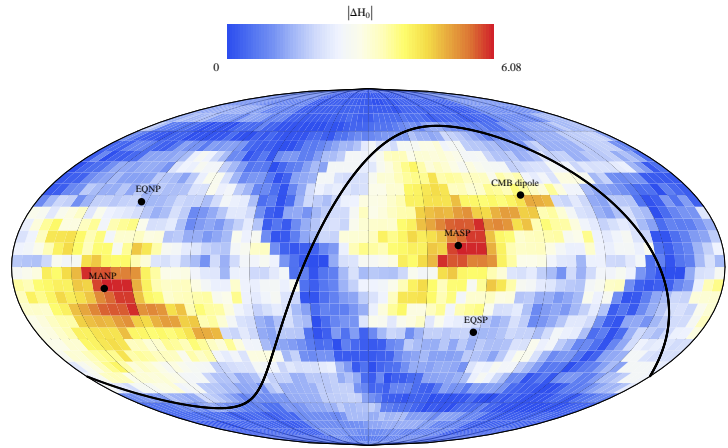,scale=0.5}
\end{minipage}
\begin{minipage}[t]{16.5 cm}
\caption{Hemispherical asymmetry of $H_0$ from the Constitution set using the SALT2 light curve 
fitter for 183 SN Ia at $z<0.2$. The plot shows $\Delta H_0 \equiv H_0(\mbox{North}) - H_0(\mbox{South})$ in units of km/s/Mpc. Thus the maximum corresponds to an asymmetry of 5\%. \label{fig1}}
\end{minipage}
\end{center}
\end{figure}

In \cite{sw07} a test was proposed to search for anisotropies of the Hubble 
diagram. The idea is to determine the best fit Hubble diagrams for opposite hemispheres. Then one can 
determine the asymmetry $|\Delta H_0|$. In \cite{sw07} we found asymmetries at the $95\%$ C.L., which 
were consistent with the assumption of unidentified systematic errors or statistical flukes. We have 
updated this analysis recently by means of the Constitution set.  

Figure 3 shows the asymmetry of a fit of the luminosity-redshift SN Ia data for all SN from the Constitution set at $z < 0.2$. Each pixel shows $\Delta H_0$ when the axis of the hemisphere is oriented towards 
the direction of the shown pixel. We fitted $H_0$ and $q_0$ for all four light-curve fitters. The asymmetry 
$(H_{\rm N} - H_{\rm S})/(H_N + H_S) \sim 5\%$ in case of the SALT light curve fitter. In contrast to the MLCS2k2 fittter, the SALT2 method leads to an isotropic $\chi^2/$d.o.f. for the quality of the fit as a function of orientation of the hemispheres. We plot galactic coordinates, the great circle is the equatorial plane.  An evaluation of the statistical significance of this finding will be presented elsewhere. So far 
we have not been able to identify any systematic issue that could explain the asymmetry. 

It should also be stressed that a $5\%$ variation of the Hubble constant corresponds to $0.1$ mag. 
The evidence for acceleration is an effect of $\sim 0.2$ mag. Thus the test of the anisotropy of the Hubble diagram is relevant for our understanding of cosmology.

\section{Conclusion}

Assuming that SN Ia are standardizable candles and that they fairly sample a 
homogeneous and isotropic Universe, the evidence for acceleration can be tested in a model-
and calibration-independent way. Various light-curve fitting procedures have been proposed 
and tested. While several fitters give consistent results for the so-called Constitution set, they lead to 
inconsistent results for the recently released SDSS SN. For MLCS2k2 with $R_V = 3.1$, SALT or 
SALT2 and relying on the Constitution set \cite{hea09}, cosmic acceleration is detected by a purely kinematic test at $> 7.6 \sigma$ when spatial flatness is assumed and at $> 4.2 \sigma$ without assumption on the spatial geometry. A weak point of the described method is the local set of SN 
(at $z < 0.2$), as these SN are essential to anchor the Hubble diagram.

One can test the local homogeneity (if we do not live at the center of an exactly spherical 
inhomogeneity) by the isotropy of the Hubble diagram. Here we presented first results from an update 
of \cite{sw07}. It seems that there is such an anisotropy at the $5\%$ level. A more detailed 
investigation will be published elsewhere. 

\section*{Acknowledgement}

DJS wishes to thank the organizers for a very interesting and inspiring school.

\end{document}